\documentclass[prl,twocolumn,showpacs,amsmath,amssymb]{revtex4-1}

\usepackage{graphicx}
\usepackage{dcolumn}
\usepackage{array}
\usepackage{bm}
\usepackage[hang]{footmisc}

\usepackage[usenames,dvipsnames]{color}

\begin{document}

\title{\bf{Effects of quantum confinement on excited state properties of SrTiO$_3$ from \textit{ab initio} many-body perturbation theory} \\[11pt] } \author{Sebastian E. Reyes-Lillo$^{(1,2)}$, Tonatiuh Rangel$^{(1,2)}$, Fabien Bruneval$^{(1,2,3)}$ and Jeffrey B. Neaton$^{(1,2,4)}$}
\affiliation{
(1) Molecular Foundry, Lawrence Berkeley National Laboratory, Berkeley, California 94720, USA \\
(2) Department of Physics, University of California, Berkeley, California 94720 USA \\
(3) CEA, DEN, Service de Recherches de M\'{e}tallurgie Physique, 91191 Gif-sur-Yvette, France\\
(4) Kavli Energy NanoSciences Institute at Berkeley, Berkeley, California, 94720 USA
}\date{\today}

\marginparwidth 2.5in
\marginparsep 0.5in
\def\seb#1{\textcolor{blue}{\small SEB: #1}}
\def\scr{\scriptsize}


\begin{abstract}
The Ruddlesden-Popper (RP) homologous series Sr$_{n+1}$Ti$_{n}$O$_{3n+1}$ provides a useful template for the study and control of the effects of dimensionality and quantum confinement on the excited state properties of the complex oxide SrTiO$_3$. We use \textit{ab initio} many-body perturbation theory within the $GW$ approximation and the Bethe-Salpeter equation approach to calculate quasiparticle energies and absorption spectrum of Sr$_{n+1}$Ti$_{n}$O$_{3n+1}$ for $n=1-5$ and $\infty$. Our computed direct and indirect optical gaps are in excellent agreement with spectroscopic measurements. The calculated optical spectra reproduce the main experimental features and reveal excitonic structure near the gap edge. We find that electron-hole interactions are important across the series, leading to significant exciton binding energies that increase for small $n$ and reach a value of 330~meV for $n=1$, a trend attributed to increased quantum confinement.  We find that the lowest-energy singlet exciton of Sr$_2$TiO$_4$ ($n=1$) localizes in the 2D plane defined by the TiO$_2$ layer, and explain the origin of its localization. 
\end{abstract}

\maketitle

Strontium titanite (SrTiO$_3$) is a prototypical ABO$_3$ perovskite that, in pure, doped or strained form, displays important dielectric, optoelectronic, and transport properties, such as ferroelectricity~\cite{Haeni2004}, robust photocatalysis~\cite{Wrighton1976}, superconductivity~\cite{Schooley1964, Kozuka2009} and high electron mobility at low temperatures~\cite{Son2010}. Atomically thin two-dimensional layers in SrTiO$_3$ heterostructures display exotic electronic effects, such as an emergent two-dimensional electron gas~\cite{Ohtomo2004, Thiel2006}, magnetism~\cite{Brinkman2007}, and metal-insulator transitions~\cite{Wang2010}. Recently, anomalous excitonic phenomena have been observed in the optical spectrum of cubic SrTiO$_3$~\cite{Sponza2013, Gogoi2015, Gogoi2016}. These effects, previously overlooked, arise due to strong Coulomb interactions between excited electrons and holes in the absence of strong screening, and can be enhanced through dimensionality and quantum confinement. Fundamental understanding of many-body electronic interactions in complex oxides and their interplay with structural degrees of freedom can lead to the design of novel functionalities and the manipulation of optical properties through electrostatic boundary conditions, size effects, or layered heterostructure. 

\begin{figure}[t]
\includegraphics[width=\columnwidth]{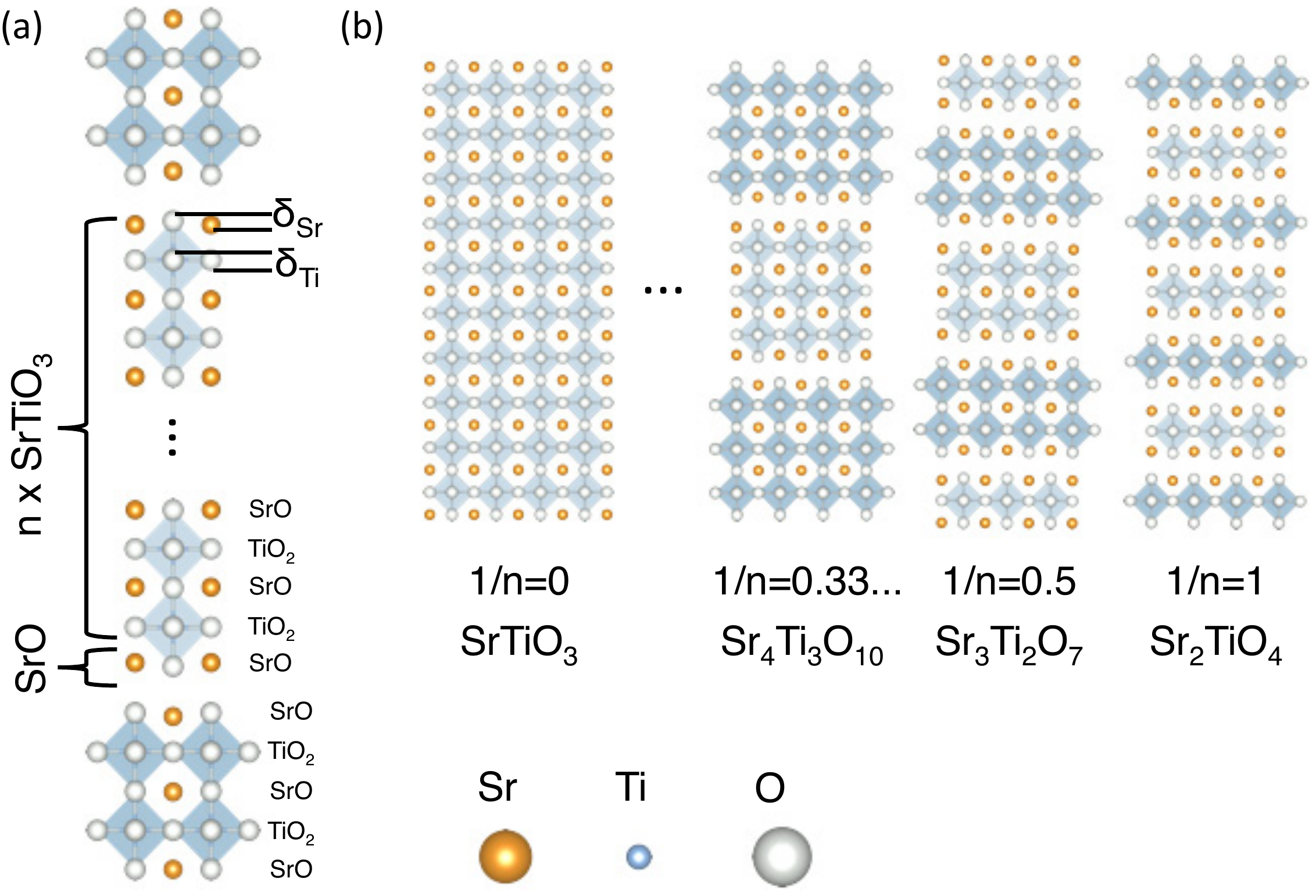}
\caption{\label{fig:fig1} (a) Structure of the $n$-th member of the RP series. The distortion $\delta_{Sr(Ti)}$ is defined as the distance between the Sr (Ti) and O atoms along the vertical direction. Sr, Ti and O atoms in orange, light-blue and white, respectively. (b) RP series as a function of $1/n$ for $n=\infty,..,3,2,1$.}
\end{figure}

The homologous Sr$_{n+1}$Ti$_{n}$O$_{3n+1}$ Ruddlesden-Popper~(RP) series is a well-studied class of materials where, via stoichiometry, one can systematically alter confinement effects on electronic structure. As shown in Fig.~\ref{fig:fig1}(a), the $n$-th member of the RP series, which takes up the tetragonal \textsl{I4/mmm} structure, consists of a periodic vertical stacking of alternating SrO and TiO$_2$ layers with an additional SrO rocksalt layer every $n$ SrTiO$_3$ perovskite unit cells along the stacking direction. As $n$ decreases, the three-dimensional corner-sharing network of octahedra of cubic SrTiO$_3$ transforms into the quasi two-dimensional layered structure of Sr$_2$TiO$_4$~($n=1$) with oxygen octahedra connected only in the $x-y$ plane (Fig.~\ref{fig:fig1}(b)); the resulting quantum well, defined by the region of connected octahedra, leads to localization due to the modification of Coulomb interactions along the vertical direction and the associated quantum confinement, with significant potential consequences for excited-state properties, as we show here.

In this work, we perform \textit{ab initio} many-body perturbation theory (MBPT) calculations on the RP series and investigate the effect of dimensionality and quantum confinement on quasiparticle energies and absorption spectrum of the complex oxide SrTiO$_3$. Several members of the RP series ($n=1-5, 10$ and $\infty$) have been experimentally stabilized through molecular beam epitaxy and under epitaxial strain~\cite{Lee2013b}. While most of the interest in the RP series has focused on its dielectric~\cite{Lee2013a} and ferroelectric properties~\cite{Lee2013a, Birol2011}, several members of the series ($n=1,2,3$) display photocatalytic activity~\cite{Jeong2006, Ko2002, Sun2015}. Additionally, recently-synthesized halide RP phases show promise in photovoltaic applications~\cite{Grote2014, Yaffe2015}. Earlier density functional theory (DFT) calculations for the RP series~\cite{Reshak2008, Lee2013b, Guan2015} captured the qualitative trend of the experimental indirect optical gaps. However, these studies neglected electron-hole interactions and excitonic effects, and proper description of quasiparticle and optical properties requires higher level of theory, as used here.

Quasiparticle and optical spectra of selected members of the Sr$_{n+1}$Ti$_{n}$O$_{3n+1}$ series are obtained with \textit{ab initio} MBPT within the $GW$ approximation~\cite{Hedin1965, Hybertsen1986} and the Bethe-Salpeter equation (BSE) approach~\cite{Rohlfing1998, Onida2002}; which has been shown to successfully describe the closely related compound TiO$_2$~\cite{Kang2010}. In this method, direct ($E^{\mathrm{dir}}_{\mathrm{el}}$) and indirect ($E^{\mathrm{ind}}_{\mathrm{el}}$) quasiparticle band gaps are computed with $GW$, where the Green's function \textit{G} and screened potential \textit{W} are constructed from mean-field energies and wave functions from DFT. Direct optical gaps ($E^{\mathrm{dir}}_{\mathrm{opt}}$) are extracted from the BSE solution through the imaginary part of the electronic dielectric function $\epsilon_2(\omega)$, using the $GW$ calculations as input. Indirect optical gaps ($E^{\mathrm{ind}}_{\mathrm{opt}}$) are estimated by subtracting the exciton binding energy ($E_{\mathrm{ex}}=E^{\mathrm{dir}}_{\mathrm{opt}}-E^{\mathrm{dir}}_{\mathrm{el}}$) from indirect $GW$ band gaps, neglecting electron-phonon effects. 

First-principles DFT calculations are performed within the local density approximation (LDA)~\cite{Perdew1981} using the \textsc{abinit} code~\cite{Gonze2009}. Our \textsc{abinit} calculations use a plane wave basis energy cut-off of~820~eV, Monkhorst-Pack \textbf{k}-point grids~\cite{Monkhorst1976}, and the recently proposed~\texttt{ONCVPSP} norm-conserving pseudo potentials~\cite{Hamann2013}, including semicore states ($4s-4p$ for Sr and $3s-3p$ for Ti) in the valence. Hybrid functional calculations, within the Heyd-Scuseria-Ernzerhof (HSE06) exchange-correlation functional~\cite{Krukau2006}, are performed with the \textsc{vasp} code~\cite{Kresse1996}. Our \textsc{vasp} calculations use an energy cut-off of 450~eV and projected augmented wave pseudopotentials~\cite{Kresse1999}. Electronic and optical properties are computed at the experimental lattice constants~\cite{Lee2013b}, but with optimized internal atomic positions. 

Single shot $GW$ ($G_0W_0$) and $G_0W_0$/BSE calculations are performed with~\textsc{BerkeleyGW} code~\cite{Deslippe2012}. We use DFT-LDA mean-field calculations as a starting point, 200 bands per atom, an screened Coulomb energy cut-off of 190~eV (54~eV) for the self-energy (electron-hole kernel), and the generalized plasmon-pole model (GPP)~\cite{Hybertsen1986} of Godby-Needs~\cite{Godby1989}. Extensive convergence studies in the number of bands and \textbf{k}-point grids show that $G_0W_0$/BSE band gaps are converged within~0.1~eV.  Further details of the calculations are included in the supplemental information (SI) \footnote{See Supplemental Material at [URL will be inserted by publisher]}.
 
Epitaxial strain has been shown to systematically modify the band gap of SrTiO$_3$~\cite{Berger2011}. In the case of the RP series, prior spectroscopic measurements were performed on thin films systems for which the in-plane lattice constant of the substrate ($a =3.87$~\AA) induces a small ($<$~1~\%) compressive (tensile) strain in the experimental (theoretical) structure of each $n$. Full structural relaxation of $n=1,2,3$ and $\infty$ with DFT-LDA underestimate experimental lattice constants of free-standing compounds by~$\sim$1\%, as is typical for the LDA (see Table I in SI). As shown in Fig.~\ref{fig:fig1}(a), the symmetry of the RP structure allows an off-centering or \emph{distortion} $\delta_{Sr(Ti)}$ of the Sr (Ti) atoms in the SrO (TiO$_2$) layers. For $n=1$, we find $\delta_{Sr}=0.04a$, in agreement with previous calculations~\cite{Noguera2000, Fennie2003} and experiments~\cite{Venkateswaran1987,Kawamura2015}.  For large $n$ ($n>4$), the distortion converges to $\delta_{Sr}=0.05a$~and $\delta_{Ti}=0.01a$ next to the SrO layer, and decreases rapidly toward the center of the perovskite layers at fixed $n$ (see Fig.~1 in SI).   

\begin{figure}[t]
\includegraphics[width=\columnwidth]{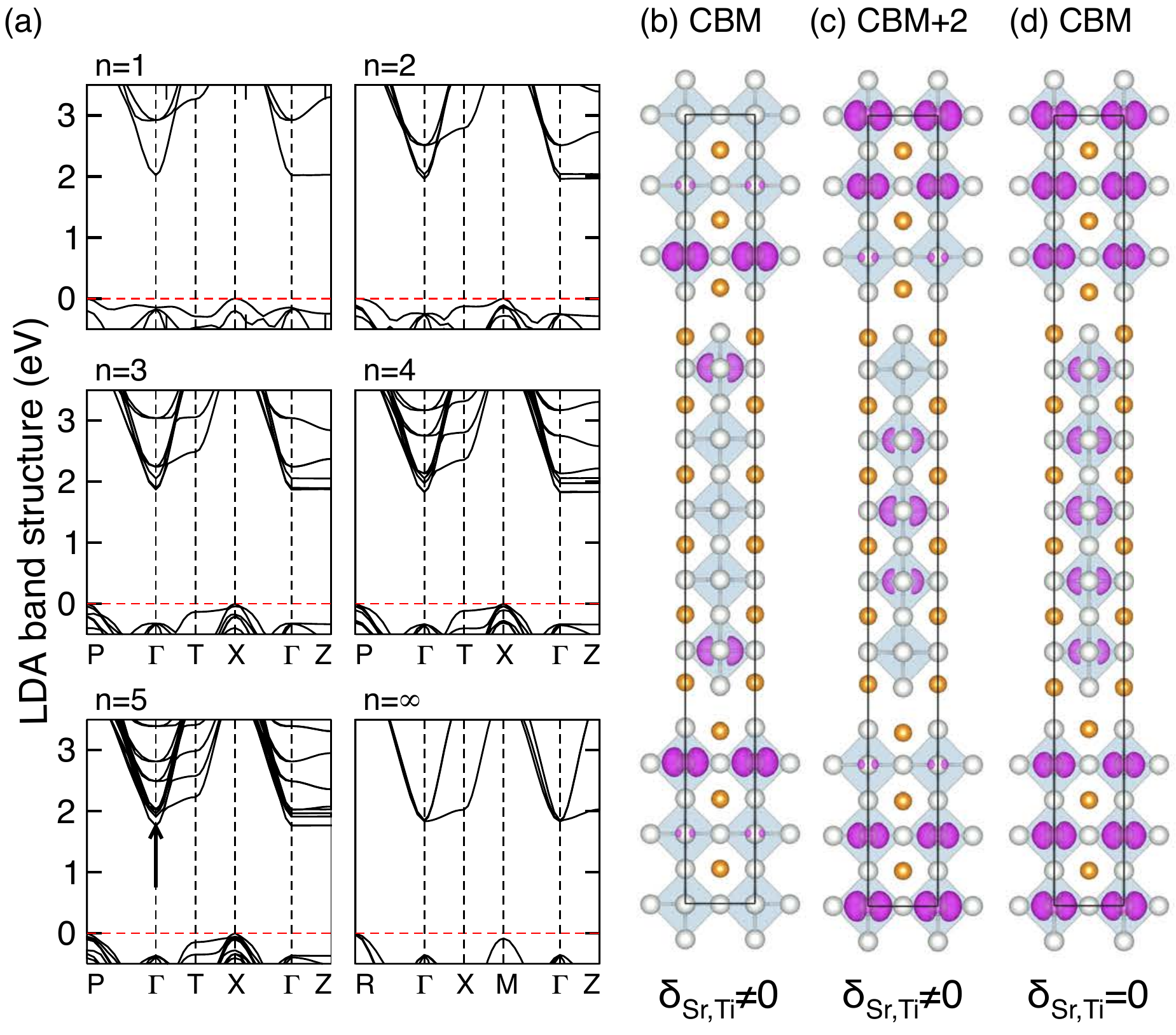}
\caption{\label{fig:fig2} (a) LDA band structure for $n=1-5$ and $n=\infty$. The arrow points to the low-lying conduction states of $n=5$. Surface of constant charge density (violet) at $\Gamma$ for (b) CBM and (c) CBM$+2$ of $n=5$, and (d) CBM of $n=5$ in the absence of distortion $\delta_{Sr,Ti}$. Atom color code is as in Fig.~\ref{fig:fig1}.}
\end{figure}

We start by considering the single-particle electronic properties of the RP series as a function of $n$. Fig.~\ref{fig:fig2}(a) shows the LDA band structure of $n=1-5$ and $\infty$ along equivalent paths in the corresponding Brillouin zones (see Table II and Fig.~2 in SI). Direct ($\Gamma$ $\rightarrow$ $\Gamma$) and indirect ($X$ $\rightarrow$ $\Gamma$ for $n<\infty$, $R$ $\rightarrow$ $\Gamma$ for $n=\infty$) band gaps are composed of transitions primarily between O $2p$ and Ti $3d$ states. For $n<\infty$, the conduction band minimum (CBM) and the valence band maximum (VBM) are degenerate: the CBM is at both $\Gamma$ and $Z$, whereas the VBM is at $P$ and $X$. The degeneracy is due to the reduced dimensionality of the system. As $n$ increases, the dispersion of conduction bands decreases along the $\Gamma-T$ symmetry line. For $n=\infty$, we recover the band structure of SrTiO$_3$.

\begin{figure}[t]
\includegraphics[width=\columnwidth]{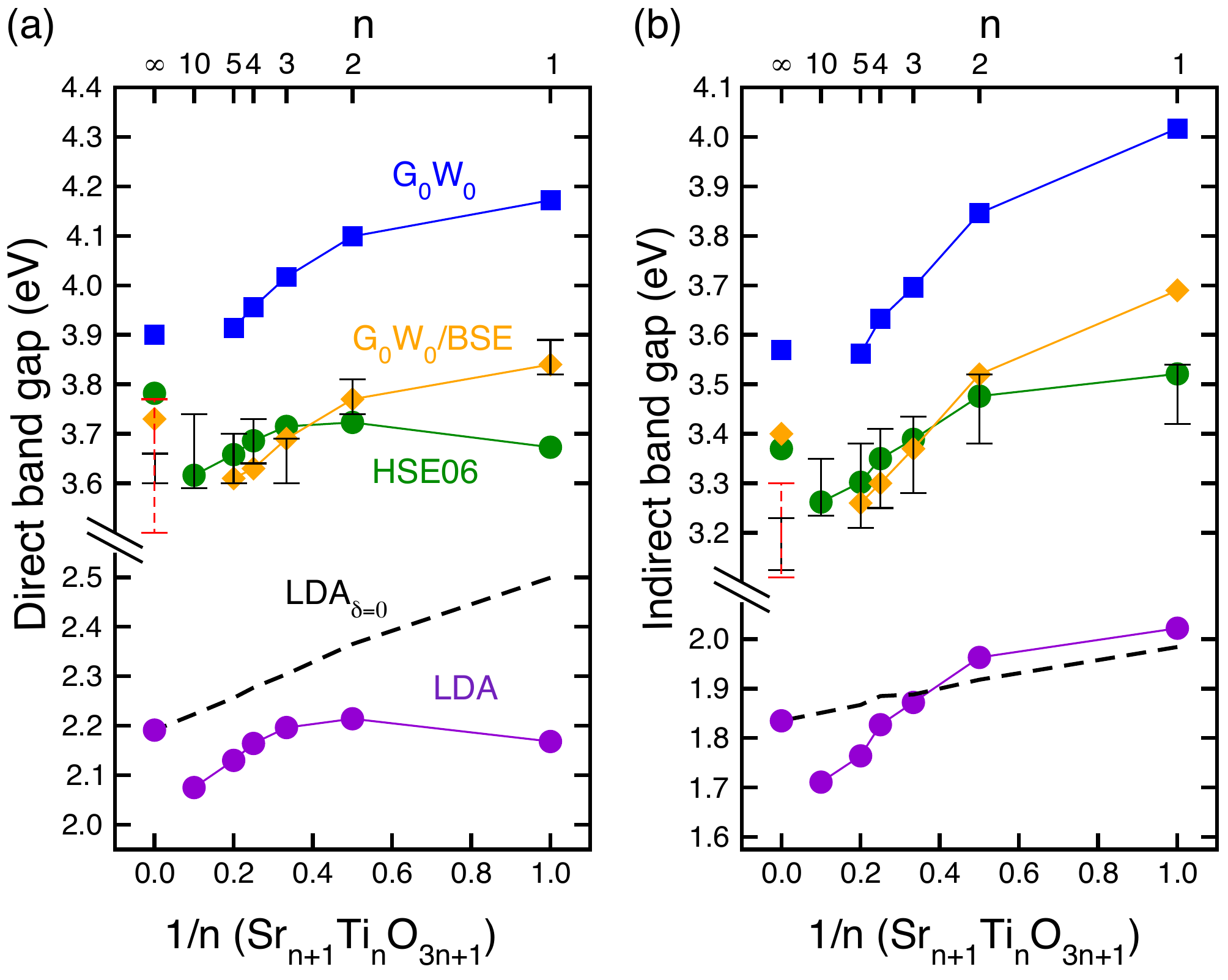}
\caption{\label{fig:fig3} (a) Direct and (b) indirect band gaps as a function of $1/n$ using different levels of theory. Experimental values are taken from Ref.~\cite{Lee2013b} (black solid error bars) and Table~\ref{table:optical} (red dash error bars).}
\end{figure}

Computed direct and indirect LDA band gaps are shown as a function of $1/n$ at the bottom of Fig.~\ref{fig:fig3}. Compared to spectroscopic measurements (error bars)~\cite{Lee2013b}, direct and indirect LDA band gaps display two important differences. First, band gaps are severely underestimated by $\sim$1.5~eV, as is usual for LDA Kohn-Sham states. Although, indirect LDA band gaps reproduce the experimental trend of indirect optical band gaps~\cite{Lee2013b}, direct LDA band gaps deviate from experimental trends at small $n$, where quantum confinement effects are expected to play an important role. Second, band gaps display a discontinuity of $\sim$0.2~eV for large $n$, which is not observed in experiments. Different choices of semi-local functionals or lattice constants (e.g. experimental, relaxed or epitaxially strained) display the same discontinuity, up to a rigid shift of~$\sim$0.1~eV. 

The band gap discontinuity for large $n$ arises from the low-lying conduction states that are significant for large finite $n$ ($n>4$) but are absent for $n=\infty$ (see Fig.~\ref{fig:fig2}(a)). These states arise due to the distortion of the two Ti atoms adjacent to the extra SrO layers and the associated splitting of the local Ti $t_{2g}$ states. As shown in Fig.~\ref{fig:fig2}(b) for $n=5$, the CBM wavefunction for large finite $n$ corresponds to two Ti $d_{xy}$ states localized at the TiO$_2$ layers next to the extra SrO layers. The bands above the low-lying conduction states (CBM$+2$) are formed by $n-2$ Ti d$_{xy}$ states that localize at the center of the perovskite layers (see Fig.~\ref{fig:fig2}(c)). The energy difference between the CBM and CBM+2 converges to $\Delta\sim0.2$~eV at large $n$ and leads to the discontinuity. 

The distortions at the extra SrO layers localize the CBM (VBM) orbital at the edges (center) of the perovskite layers, leading to the quasi two-dimensional properties of the RP series. If the distortions are removed from the structure of $n$ by artificially aligning the Sr and Ti atoms with the O atoms, the CBM becomes a degenerate set of $n$ Ti $d_{xy}$ states and the CBM wavefunction recovers to an unconfined uniform distribution of $d_{xy}$ states along the vertical direction (see Fig.~\ref{fig:fig2}(d)). Note that if the distortions are removed from the entire series, LDA band gaps display a linear trend as a function of $1/n$ that deviates from experiments at large $n$ (see LDA$_{\delta=0}$ in Fig.~\ref{fig:fig3}).

Before discussing our \textit{ab initio} MBPT results, we compare our LDA results with hybrid functional calculations. As a function of $1/n$ (Fig.~\ref{fig:fig3}), HSE06 band gaps are in good agreement with measured optical band gaps. However, this agreement, while good, is fortuitous, since the generalized Kohn-Sham system is formally unable to describe interacting two-particle states, e.g. excitons. Trends in HSE06 band gaps are well approximated by a rigid shift of LDA band gaps, and in particular, underestimate by $\sim0.2$~eV the experimental direct optical band gap of $n=1$,  where electron-electron and electron-hole interaction effects are expected to be important.

\begin{table}[b]
\caption{Electronic quasiparticle band gaps (eV) for cubic SrTiO$_3$ and Sr$_2$TiO$_4$ using different levels of theory.}
\begin{ruledtabular}
\begin{tabular}{lcc}

\begin{tabular}{l}  
\\    \\   This work\\ $G_0W_0$(GPP)/LDA\\ $G_0W_0$(FF)/LDA\\  Prev. work\\  $G_0W_0$/LDA\\ $G_0W_0$/LDA\\  $G_0W_0$/LDA+U\\   $G_0W_0$/LDA\\ Expts.\\
\end{tabular}

\begin{tabular}{cc}
\begin{tabular}{c}
SrTiO$_3$ \\ \hline
\begin{tabular}{cc}
\begin{tabular}{c} $E^{\mathrm{dir}}_{\mathrm{el}}$\\   \\   3.90\\  3.86\\   \\ 3.76 \\ 3.80\\ 3.77\\      \\   3.30\\  \end{tabular}
&\begin{tabular}{c} $E^{\mathrm{ind}}_{\mathrm{el}}$\\   \\   3.57\\ 3.52\\  \\        \\ 3.40\\ 3.28\\   3.32\\  \\ \end{tabular}
\end{tabular}
\end{tabular}
&
\begin{tabular}{c}
Sr$_2$TiO$_4$ \\ \hline
\begin{tabular}{cc}
\begin{tabular}{c} $E^{\mathrm{dir}}_{\mathrm{el}}$\\    \\  4.17\\ 4.21\\  \\ \\ \\ \\ \\  \\ \end{tabular}
&\begin{tabular}{c} $E^{\mathrm{ind}}_{\mathrm{el}}$\\   \\   4.02\\ 4.06\\ \\ \\ \\ \\ \\ \\  \end{tabular}
\end{tabular}
\end{tabular}
\end{tabular} 

\begin{tabular}{l}  
Ref.\\  \\  \\  \\  \\  \\ \cite{Sponza2013, Gogoi2015}\\ \cite{Gogoi2016}\\ \cite{Kim2010}\\  \cite{Hamann2009}\\ \cite{Tezuka1994} \\
\end{tabular}

\end{tabular}
\end{ruledtabular}
\label{table:electronic}
\end{table}

We now turn to our results with \textit{ab initio} MBPT.  Table~\ref{table:electronic} shows fundamental band gaps for cubic SrTiO$_3$ and $n=1$ calculated with $G_0W_0$/LDA. Our $G_0W_0$ calculations for cubic SrTiO$_3$ are in line with previous $GW$ results (by $\sim0.1$~eV) but overestimate photoemission measurements (by $\sim0.6$~eV). Full-frequency $GW$(FF) calculations for cubic SrTiO$_3$ and $n=1$ are well approximated by $GW$(GPP), and we therefore use the GPP for the rest of the series. As a function of $1/n$ (Fig.~\ref{fig:fig3}), direct (indirect) $G_0W_0$ band gaps increase monotonically by $\sim$0.3~eV (0.5~eV), with a steeper slope than HSE06 band gaps. $G_0W_0$ band gaps overestimate experimental optical band gaps due to the absence of electron-hole interactions. For $n=1$, the measured direct (indirect) optical gap is overestimated by $\sim$0.3~eV (0.5~eV).

\begin{table}[b]
\caption{Optical gaps (eV) for cubic SrTiO$_3$ and Sr$_2$TiO$_4$ from an exact diagonalization solution of the BSE.}
\begin{ruledtabular}
\begin{tabular}{lcc}

\begin{tabular}{l}  
 \\ \\  This work\\ $G_0W_0$/BSE\\ Prev. work\\ $G_0W_0$/BSE\\ Expts.\\ \\ \\ \\ \\ \\ \\
\end{tabular}

\begin{tabular}{cc}
\begin{tabular}{c}
SrTiO$_3$ \\ \hline
\begin{tabular}{cc}
\begin{tabular}{c} $E^{\mathrm{dir}}_{\mathrm{opt}}$\\      \\  3.73\\  \\ 3.54\\ 3.80\\ 3.63\\         \\ 3.75\\                \\ \\ 3.90\\  \end{tabular}
&\begin{tabular}{c} $E^{\mathrm{ind}}_{\mathrm{opt}}$\\   \\ 3.40\\   \\  \\  3.20\\ 3.15\\ 3.20\\  3.25\\         3.10\\    3.29\\   \\ \end{tabular}
\end{tabular}
\end{tabular}
&
\begin{tabular}{c}
Sr$_2$TiO$_4$ \\ \hline
\begin{tabular}{cc}
\begin{tabular}{c} $E^{\mathrm{dir}}_{\mathrm{opt}}$\\      \\ 3.84\\   \\  \\ \\ 3.86\\ \\ \\          \\   \\  \\ \end{tabular}
&\begin{tabular}{c} $E^{\mathrm{ind}}_{\mathrm{opt}}$\\   \\ 3.69\\   \\ \\ \\ 3.48\\ \\ \\   3.40\\   \\ \\ \end{tabular}
\end{tabular}
\end{tabular}

\end{tabular} 

\begin{tabular}{l}  
Ref.\\ \\   \\ \\  \\ \cite{Sponza2013, Gogoi2015}\\  \cite{Gogoi2016}\\ \cite{Lee2013b}\\  \cite{Cardona1965}\\  \cite{Benthem2001}\\   \cite{Eng2003}\\  \cite{Dejneka2010}\\ \cite{Matsuno2005}\\
\end{tabular}

\end{tabular}
\end{ruledtabular}
\label{table:optical}
\end{table}


Table~\ref{table:optical} shows $G_0W_0$/BSE direct optical gaps for cubic SrTiO$_3$ and $n=1$ as obtained from an exact diagonalization solution of the BSE. These calculations correctly include the electron-hole attraction that reduces the optical gap relative to the quasiparticle (fundamental) band gap. Our direct BSE optical gaps for SrTiO$_3$ are in good agreement with previous calculations and experiments. The large overestimation of $G_0W_0$ for $n=1$ is compensated by a large increase of exciton binding energy, which almost doubles from $n=\infty$ (170~meV) to $n=1$ (330~meV), an increase that can be attributed to quantum confinement. Note that the increase of exciton binding energy is not explained by trends in effective mass and electronic dielectric constant as a function of $n$ (see SI).

As a function of $1/n$ (Fig.~\ref{fig:fig3}), $G_0W_0$/BSE band gaps extracted from the first feature (peak or shoulder for $n<\infty$, first excitation for $n=\infty$) of the $G_0W_0$/BSE optical spectrum are in excellent agreement (by $\sim0.1$~eV) with measured optical gaps. The best agreement is obtained for small $n$ direct gaps, where indirect phonon-assisted transitions are irrelevant. We find relatively large electron-hole interaction across the entire series, the exciton binding energy remains large for several members of the series: $n=1-4$  (330 meV) and $n=5$ (300 meV), since the CBM remains localized at the TiO$_2$ layers next to the extra SrO layers.

\begin{figure}[t]
\includegraphics[width=\columnwidth]{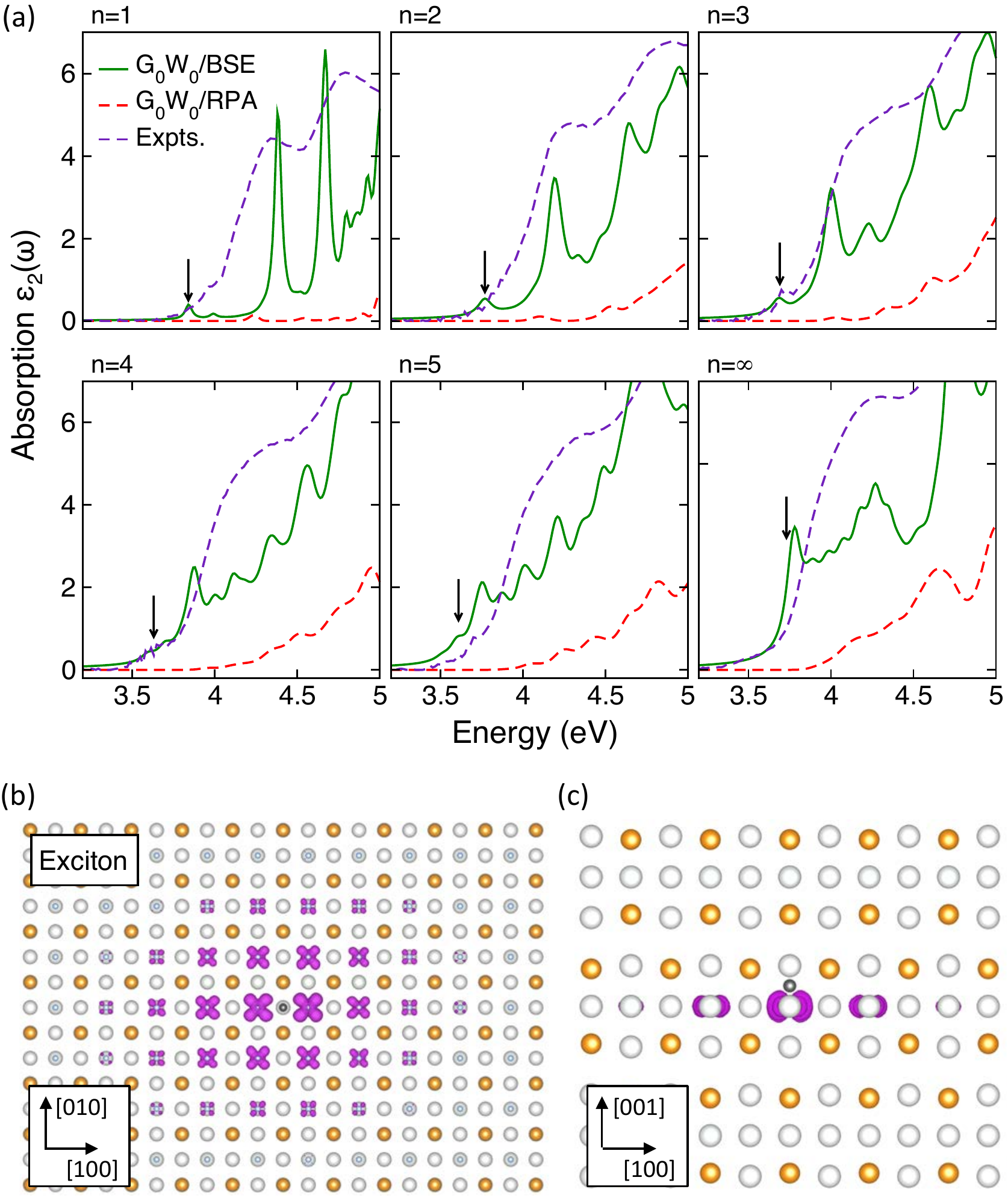}
\caption{\label{fig:fig4} (a) Absorption spectra for Sr$_{n+1}$Ti$_n$O$_{3n+1}$ ($n=1-5$ and $\infty$) with (solid green) and without (dashed red) electron-hole interactions. The black arrows show the BSE band gap. Experiments (dashed violet) are taken from Ref.~\cite{Lee2013b}. (b) Top and (c) side view of the exciton wave function of $n=1$ (violet) and the position of the hole in the oxygen atom (black dot).}
\end{figure}

We focus now in the properties of the $G_0W_0$/BSE spectrum as a function of $n$. Fig.~\ref{fig:fig4}(a) shows the experimental and computed absorption spectra for $n=1-5$ and $\infty$ with ($G_0W_0$/BSE) and without ($G_0W_0$/RPA) the inclusion of electron-hole interactions. The $G_0W_0$/BSE spectra correctly reproduces the position and magnitude of the experimental onset and reveal excitonic structure. The spectrum of $n=1$ displays two-dimensional characteristics and strong excitonic peaks; there is a first peak at 3.84~eV, a weak second peak at 4.00~eV, and a strong third peak at 4.40~eV. From direct examination of the lowest-energy BSE solution, the first peak arises primarily from direct transitions to the Ti $d_{xy}$ conduction band. The large energy difference (0.5~eV) between the dominant first and third peak is explained by the large energy difference (0.9~eV) between the CBM and the rest of the conduction bands. The first peak merges with the rest of the spectrum at a critical layer thickness $n_c=4$, and the third peak becomes the spectrum onset at $n=\infty$. 

As $n$ increases, there is a large transfer of spectral weight toward small energies due to the decrease of dispersion along the $\Gamma-T$ line (see Fig.~\ref{fig:fig2}~(a)). At the critical layer thickness $n_c$, transitions to the CBM at the SrO layers become negligible compared to transitions to the CBM$+2$ at the center of the perovskite layers. As already suggested in Ref.~\cite{Lee2013b}, the volume of the increasingly large region of connected octahedra dominates over the thin SrO extra layers, and the spectrum approaches the spectrum of $n=\infty$. Experimentally, low-lying conduction states and excitonic effects become negligible due to limited resolution (see Fig.~\ref{fig:fig4}~(a)) and optical gaps converge continuously with increasing $n$, explaining the absence of band gap discontinuity in the optical measurements (see Fig.~\ref{fig:fig3}~(a)).

The increase of exciton binding energy at small $n$ implies a spatial localization of the exciton of $n=1$. As a means of visualizing the lowest singlet exciton, Fig.~\ref{fig:fig4}(b)-(c) shows the top and side views of an isosurface probability amplitude of finding an electron given a hole is fixed near the O atom. The exciton displays charge-transfer character and localizes, in the 2D plane defined by the lone TiO$_2$ layer. The real space wave function envelope of the exciton resembles the $1s$ state of the 2D hydrogenic model and has an average radius of $R\sim 5 a_B$~\cite{Sharifzadeh2013a}, with $a_B$ the Bohr radius. Our calculations show that the exciton binding energy is insensitive to epitaxial strain: while direct and indirect band gaps decrease linearly by~$\sim0.5$~eV from tensile ($-4$\%) to compressive ($+4$\%) strain, the exciton binding energy varies by less than 10~meV in the same strain range. As $n$ increases, the exciton localizes on the TiO$_2$ layers next to the extra SrO layer due to the distortions next to the SrO layers. 

In summary, we have used DFT with different functionals, and the $G_0W_0$/BSE approach to compute quasiparticle and optical spectra of selected members of the Sr$_{n+1}$Ti$_{n}$O$_{3n+1}$ series. Our $G_0W_0$/BSE direct and indirect optical gaps for the RP series are in excellent agreement with measured optical gaps when electron-hole interactions are included. We find that quantum confinement, induced by the extra SrO layers in the RP series, produces a larger increase in quasiparticle band gaps, relative to trends in optical gaps. The exciton binding energy is significant and increases for small $n$, reaching a value of 330~meV. The $G_0W_0$/BSE spectra reproduce the main experimental features and reveal excitonic structure near the gap edge. Our work suggests that the RP series may be a useful framework for the study of dimensionality and quantum confinement in optoelectronic properties of complex oxides. 
\medskip

\section{\label{sec:level1} Acknowledgments}

S.E.R.-L. thanks R.F. Berger and T. Birol for valuable discussions. Work at the Molecular Foundry was supported by the Office of Science, Office of Basic Energy Sciences, of the U.S. Department of Energy, and Laboratory Directed Research and Development Program at the Lawrence Berkeley National Laboratory under Contract No. DE-AC02-05CH11231. T.R. acknowledges support from the SciDAC Program on Excited State Phenomena in Energy Materials. F.B. acknowledges the Enhanced Eurotalent program and the France Berkeley Fund for supporting his sabbatical leave in UC Berkeley. This research used resources of the National Energy Research Scientific Computing Center, which is supported by the Office of Science of the U.S. Department of Energy.

\end{document}